\documentclass[12pt]{iopart}
\begin{document}

\def\beq{\begin{eqnarray}}
\def\eeq{\end{eqnarray}}
\def\s{\mbox{\boldmath$\displaystyle\mathbf{\sigma}$}}
\def\bp{\mbox{\boldmath$\displaystyle\mathbf{\pi}$}}
\def\be{\mbox{\boldmath$\displaystyle\mathbf{\eta}$}}
\def\J{\mbox{\boldmath$\displaystyle\mathbf{J}$}}
\def\K{\mbox{\boldmath$\displaystyle\mathbf{K}$}}
\def\P{\mbox{\boldmath$\displaystyle\mathbf{P}$}}
\def\p{\mbox{\boldmath$\displaystyle\mathbf{p}$}}
\def\hp{\mbox{\boldmath$\displaystyle\mathbf{\widehat{\p}}$}}
\def\x{\mbox{\boldmath$\displaystyle\mathbf{x}$}}
\def\0{\mbox{\boldmath$\displaystyle\mathbf{0}$}}
\def\bv{\mbox{\boldmath$\displaystyle\mathbf{\varphi}$}}
\def\bx{\mbox{\boldmath$\displaystyle\mathbf{\xi}$}}
\def\bs{\mbox{\boldmath$\displaystyle\mathbf{\sigma}$}}
\def\bc{\mbox{\boldmath$\displaystyle\mathbf{\chi}$}}

\def\hbv{\mbox{\boldmath$\displaystyle\mathbf{\widehat\varphi}$}}
\def\hbxi{\mbox{\boldmath$\displaystyle\mathbf{\widehat\xi}$}}
\def\bn{\mbox{\boldmath$\displaystyle\mathbf{\nabla}$}}
\def\bl{\mbox{\boldmath$\displaystyle\mathbf{\lambda}$}}
\def\br{\mbox{\boldmath$\displaystyle\mathbf{\rho}$}}
\def\1{1}
\def\ar{\stackrel{\hspace{0.04truecm}grav. }{\mbox{$\longrightarrow$}}}

\letter{}

\title[Operational indistinguishabilty of doubly
special relativities]{Operational indistinguishabilty of doubly
special relativities from special relativity}

\author{D V Ahluwalia-Khalilova}  

\address{
Inter-University Centre for Astronomy and Astrophysics (IUCAA),
Post Bag 4, Ganeshkhind, Pune 411 007, India\footnote{Present address:
Department of Mathematics, University of Zacatecas,
Zacateacs, ZAC 98060, Mexico}}

\eads{d.v.ahluwalia-khalilova@heritage.reduaz.mx }

\begin{abstract}\\
We argue that existing doubly special relativities may
not be operationally distinguishable from the special
relativity.
In the process we point out that some of the  phenomenologically
motivated modifications of dispersion relations, and 
arrived conclusions, must be reconsidered. Finally, we reflect 
on the possible conceptual issues that arise in quest for
a theory of spacetime with two invariant scales. 
\end{abstract}


\maketitle

\begin{quote}{\it
This paper is no longer being pursued for publication.
Its essential, and updated, content is now available in
Sec. 2 of D. V. Ahluwalia-Khalilova, ``A freely falling
frame at the interface of gravitational and quantum realms,''
Class. Quant. Grav. {\bf 22} (2005) 1433-1450.}
\end{quote}



\noindent

\noindent
\textit{Introduction.}\textemdash~
Postponing  discussion on
the  necessity for relativities with two invariant scales
to the concluding section, we first consider recent 
proposals for such theories. They have been 
studied under the term doubly special relativity 
\cite{dsr1,gac_review}.
We show that
these suggestions, despite appearances to the contrary,  
are operationally indistinguishable from the special relativity.

The conclusion is arrived in the following fashion:
Since the effect of modified dispersion relations
is most dramatically apparent in the kinematical 
equations of motion we first look at the kinematical description of
fermions and  bosons in doubly special 
relativities. At this stage of the argument it would seem  
that the theory has 
potentially significant phenomenological implications. But,
when we transform the formalism to the recently suggested Judes-Visser
variables,  we arrive at the conclusion 
that questions 
operational 
distinguishability of these theories  from  the special 
relativity. \\

\noindent
\textit{Doubly Special Relativities of Amelino-Camelia, and
Magueijo and Smolin.}\textemdash~
Simplest of  doubly special relativities (\texttt{DSR})\footnote[1]{The
phrase ``doubly special relativity'' is somewhat confusing. The 
special 
of ``special relativity'' refers to the circumstance that 
one restricts to a special class of inertial observers which move
with  relative uniform velocity. The general of ``general relativity''  
lifts this restriction. The ``special'' of special relativity
has nothing to do with one versus two
invariant scales. It rather refers to the special class of inertial 
observers; a circumstance that remains unchanged in  special relativity
with two invariant scales. The theory of general relativity with
two invariant scales would thus not be called ``doubly general relativity.''
However, given widespread usage of the term ``doubly special relativity''
we shall use it here but with the explicit understanding
that by it one means a special relativity with two invariant scales.} 
result from
keeping the algebra of boost- and 
rotation- generators
intact while modifying the boost parameter in a non-linear manner. 
Specifically, in the \texttt{DSR} of Amelino-Camelia
the boost parameter, $\bv$, changes from the special relativistic form
\beq
\cosh{\varphi} = \frac{E}{m }\,,\quad
\sinh{\varphi} = \frac{p}{m}\,,\quad
\hbv=\frac{\p}{p}\,, \label{dirac}
\eeq
to \cite{dsr1,gak,jv}

\beq
\cosh\xi &=& 
\frac{1}{\mu}\left( \frac{
e^{\ell_P  E} 
-\cosh\left(\ell_P\, m \right)}
{\ell_P \cosh\left( \ell_P \,m/2\right)}
\right)\,,\label{gac1}\\
\sinh\xi &=& 
\frac{1}{\mu}\left(
\frac{p \,
e^{\ell_P  E}
}
{\cosh\left(\ell_P \,m/2\right)} \right) \,,
\quad
\hbxi=\frac{\p}{p}\,, \label{gac2}
\eeq
while for the \texttt{DSR}  of  Magueijo and Smolin the change takes
the form \cite{dsr2,jv}
\beq
\cosh\xi &=& 
\frac{1}{\mu}
\left(\frac{E}{1-\ell_P\,E}\right)
\,,\label{ms1}\\
\sinh\xi &=& 
\frac{1}{\mu}
\left(\frac{p}{1-\ell_P\,E}\right)\,,
\quad
\hbxi=\frac{\p}{p}
\,. \label{ms2}
\eeq
Here,  $\mu$ is a Casimir invariant of \texttt{DSR} (see Eq. 
(\ref{ci}) below) and is given by
\beq
\mu =\cases{
\frac{2}{\ell_P}\,
\sinh\left(\frac{\ell_P \,m}{2}\right) 
                     & $\mbox{for  Ref. \cite{dsr1}'s \texttt{DSR} }$\\
\frac{m}{1-\ell_P m}
                     & $\mbox{for  Ref. \cite{dsr2}'s \texttt{DSR} }$\cr}
\eeq
The notation is that of Ref. \cite{jv}; with the minor  exceptions:
$\lambda$, $\mu_0$, $m_0$ there are $\ell_P$,  $\mu$, $m$ here.

Now,  it is an assumption of \texttt{DSR} theories that 
the non-linear action of $\bx$ is restricted to the momentum space 
\textit{only}. No fully satisfactory spacetime description
in the context of the DSR theories has yet emerged, and 
we are not sure if such an operationally meaningful 
description indeed exists. Therefore, our arguments shall be made 
entirely in the momentum space. \\

\noindent
\textit{Master equation for spin-1/2: Dirac case.}\textemdash~
Since the underlying 
spacetime symmetry generators remain unchanged much of the  
formal apparatus of the finite dimensional representation
spaces associated with the Lorentz group remains intact.
In particular, there still exist $(1/2,\,0)$ and $(0,\,1/2)$
spinors. But now they transform from the rest frame to
an inertial frame in which the particle has momentum, $\p$, as:
\beq
\phi_{(1/2,\,0)}\left(\p\right)
& = & \exp\left( + \frac{\bs}{2}\cdot\bx \right)
\phi_{(1/2,0)}\left(\0\right)\,,\label{a}\\
\phi_{(0,\,1/2)}\left(\p\right)
& = & \exp\left(- \frac{\bs}{2}\cdot\bx \right)
\phi_{(0,1/2)}\left(\0\right)\,.\label{b}
\eeq  
Since 
the null momentum vector $\0$ is still isotropic,
one may assume that (see p. 44  of 
Ref. \cite{Ryder} \textit{and} Refs. \cite{dva_review,ak,gg}): 
\beq
\phi_{(0,1/2)}\left(\0\right) = \zeta \,\phi_{(1/2,0)}\left(\0\right)\,,
\label{c}
\eeq
where $\zeta$ is an undetermined phase factor. 
In general, the phase
$\zeta$ encodes
C, P, and T properties. The interplay of
Eqs. (\ref{a}-\ref{b}) and (\ref{c}) yields the Master equation for
the $(1/2,\,0)\oplus(0,\,1/2)$ spinors,
\beq
\psi\left(\p\right) = \left(
			\begin{array}{c}
			\phi_{(1/2,\,0)}\left(\p\right)\\
			\phi_{(0,\,1/2)}\left(\p\right)
			\end{array}
		     \right)\,,
\eeq
to be
\beq
\left(
\begin{array}{cc}
-\zeta \1_2 & \exp\left(\bs\cdot\bx\right) \\
\exp\left(- \bs\cdot\bx\right) & - \zeta^{-1} \1_2
\end{array}
\right) \psi\left(\p\right) = 0\,,\label{meq}
\eeq
where $\1_n$ stands for $n\times n$ identity matrix
(and $0_n$  represents the corresponding null matrix).
This is one of the central results on which would be anchored, as we would
see latter, the thesis summarized in the Abstract.

As a check, taking $\bx$ to be $\bv$, and after some
simple algebraic manipulations,
the Master equation (\ref{meq}) reduces to:
\beq
\left(
 	\begin{array}{cc}
	- m \zeta \1_2 & E \1_2 + \bs\cdot \p \\
	E \1_2 - \bs\cdot \p & - m \zeta^{-1} \1_2
	\end{array} 
\right) \,
\psi\left(\p\right) = 0\,.\label{d}
\eeq
With the given identification of the boost parameter
we are in the realm
of special relativity. There, the operation of
parity is well understood. Demanding  parity
covariance for Eq. (\ref{d}), we obtain
$\zeta=\pm 1$. Identifying 
\beq
\left(  \begin{array}{cc}
	0_2 & \1_2 \\
	\1_2 & 0_2
	\end{array}
\right)\,,\quad
\left(  \begin{array}{cc}
	0_2 & -\bs\\
	\bs & 0_2
	\end{array}
\right)\,,
\eeq
with the Weyl-representation $\gamma^0$, and $\gamma^i$, respectively;
Eq. (\ref{d}) reduces to the Dirac equation of  special relativity,
\beq
\left(\gamma^\mu p_\mu \mp m\right)\psi\left(\p\right)=0\,.\label{de}
\eeq
The linearity of the Dirac equation in $p_\mu= (E,-\p)$, is now clearly
seen to be associated with two observations: 

\begin{enumerate}
\item[$\mathcal{O}_1$.]
That, $\bs^2 = \1_2$; and
\item[$\mathcal{O}_2$.]
That in special relativity, the hyperbolic functions
\textendash~ 
see Eq. (\ref{dirac}) \textendash~ associated with the boost parameter
are linear in $p_\mu$. 
\end{enumerate}
In \texttt{DSR}, observation $\mathcal{O}_1$
still holds. But, as Eqs. (\ref{gac1} - \ref{ms2}) 
show, $\mathcal{O}_2$ is strongly violated.
The extension of the presented formalism for Majorana spinors 
is more subtle \cite{m1,m2,m3} and can be carried using the
techniques developed in Ref. \cite{hep-ph/0212222}.\\

\noindent
\textit{Master equation for higher spins.}\textemdash~
The above-outlined procedure applies
to all, bosonic as well as fermionic,  
$(j,0)\oplus(0,j)$ representation spaces. It is 
not confined to $j=1/2$. A straightforward generalization 
of the $j=1/2$ analysis immediately yields the Master equation 
for an arbitrary-spin,
\beq
\left(
\begin{array}{cc}
-\zeta\, \1_{2j+1}& \exp\left(2\J\cdot\bx\right) \\
\exp\left(- 2\J\cdot\bx\right) & - \zeta^{-1} \, \1_{2j+1}
\end{array}
\right) \psi\left(\p\right) = 0\,,\label{j}
\eeq
where
\beq
\psi\left(\p\right) = \left(
			\begin{array}{c}
			\phi_{(j,\,0)}\left(\p\right)\\
			\phi_{(0,\,j)}\left(\p\right)
			\end{array}
		     \right)\,.\label{js}
\eeq
Equation (\ref{j}) 
contains the central result of the previous section as a 
special case.  
For studying the special relativistic limit it is convenient 
to bifurcate the
$(j,0)\oplus(0,j)$ space into two sectors by splitting the 
$2(2j+1)$ phases, $\zeta$,
into two sets: $(2j+1)$ phases $\zeta_+$, and the other  
$(2j+1)$ phases $\zeta_-$. Then  in particle's rest frame
the $\psi(\p)$ may be written as:
\beq
\psi_h(\0)=
\cases{
u_h(\0) &  $\mbox{when}~ \zeta=\zeta_+$\cr
v_h(\0) &  $\mbox{when} ~\zeta=\zeta_- $\cr
}
\eeq
The explicit forms of $u_h(\0)$ and $u_h(\0)$ (see Eq. (\ref{c})) are:
\beq
u_h(0)=\left(
\begin{array}{c}
\phi_h(\0) \\
\zeta_+ \,\phi_h(\0)   
\end{array}
\right),\,
v_h(0)=
\left(
\begin{array}{c}
\phi_h(\0) \\
\zeta_-\, \phi_h(\0)   
\end{array}
\right),
\eeq
where the   $\phi_h(\0)$ are defined as:
$\J\cdot \hp\, \phi_h(\0) = h \,\phi_h(\0)$, and $h=-j,-j+1,\ldots,+j$.
In the parity covariant special relativistic 
limit, we find $\zeta_+ = +1$ while   
$\zeta_- = -1$.

As a check, for $j=1$, identification of $\bx$ with $\bv$,
and after implementing parity covariance, Eq. (\ref{j}) yields
\beq
\left(\gamma^{\mu\nu}p_\mu p_\nu \mp m^2\right)\psi(\p)=0\,.\label{bwweq}
\eeq 
The  $\gamma^{\mu\nu}$ are unitarily equivalent  
to those of Ref. \cite{bww}, and thus we reproduce 
\textit{bosonic matter fields}
with $\left\{C,\,P\right\} = 0$.  A carefully taken massless limit then shows
that the resulting equation is consistent with the free Maxwell equations
of electrodynamics.

Since the $j=1/2$ and $j=1$ representation spaces of \texttt{DSR} reduce to
the Dirac and Maxwell descriptions, it would seem apparent
(and as is
often argued in similar contexts \cite{aa}
\textendash~ wrongly as we will soon see \textendash~)  that
the \texttt{DSR}
contains physics beyond the linear-group realizations of special relativity.
To the lowest order in $\ell_P$,  Eq. (\ref{meq}) yields
\beq
\left(
\gamma^\mu {p}_\mu + \tilde{m} + 
\delta_1\, \ell_P\right)
\psi(\p)=0\,,
\eeq
where
\beq
\tilde{m}
&=& \left(
\begin{array}{cc}
-\zeta \1_2  & 0_2  \\
0_2 & -\zeta^{-1} \1_2
\end{array}
\right)\,m \,
\eeq
and
\beq
\delta_1 =
\cases{
\gamma^0\left(\frac{E^2-m^2}{2}\right)+\gamma^i p_i\, E
& $\mbox{for  Ref. \cite{dsr1}'s \texttt{DSR}}$\\
\gamma^\mu p_\mu\, \left(E-m\right)
&       $\mbox{for Ref. \cite{dsr2}'s \texttt{DSR}}$
}
\eeq
Similarly, the presented Master equation can be used  
to obtain \texttt{DSR}'s counterparts for Maxwell's electrodynamic.
Unlike the Coleman-Glashow framework \cite{cg}, the existing DSRs 
provide \textit{all} 
corrections, say, to the standard model of the high energy physics, 
in terms of \textit{one} \textendash~ and \textit{not forty six} \textendash~ 
fundamental constant, $\ell_P$.
Had DSRs been operationally distinct this would have been a
a remarkable power of DSR-motivated frameworks. 
\\

\noindent
\textit{Challenging the DSR's operational 
distinguishability from Special Relativistic framework:
Spin-1/2 and Spin-1 description in Judes-Visser 
Variables.}\textemdash~
We now show that the  DSR program as implemented currently is 
misleading. The question is what are the operationally measurable
quantities in DSR? The $E$ is no longer the $0$th component, 
nor is $\p$ the spatial component  of $4-$momentum. 
Neither  is $m$   an invariant under the DSR boosts. Their physical 
counterparts, as we interpret them, are  
Judes-Visser  variables \cite{jv}, $\eta^\mu \equiv 
\left(\epsilon(E,p),\,\bp(E,p)\right)=(\eta^0,\be)$, and $\mu$.
The $\epsilon(E,p)$ and $\bp(E,p)$ 
relate to the rapidity parameter $\bx$ of  \texttt{DSR} 
in same functional form
as do $E$ and $\p$ to  $\bv$ of special relativity:
\beq
\cosh\left(\xi\right)= \frac{\epsilon(E,p)}{\mu}\,,\quad
\sinh\left(\xi\right)= \frac{\pi(E,p)}{\mu}\,,
\eeq
where 
\beq
\mu^2= \left[\epsilon(E,p)\right]^2 - \left[\bp(E,p)\right]^2\,.
\label{ci}
\eeq
They provide 
the most economical and physically transparent formalism for 
representation space theory in \texttt{DSR}.
For $j=1/2$ and $j=1$, Eq. (\ref{j}) yields the \textit{exact
DSR} equations for $\psi(\bp)$:
\beq
\left(\gamma^\mu \eta_\mu + \tilde{\mu}\right)
\psi\left(\bp\right)=0\,, \quad{\mbox{where}}\;\;
\tilde{\mu}
= \left(
\begin{array}{cc}
-\zeta^{-1} \1_2 & 0_2 \\
0_2 & -\zeta \1_2
\end{array}
\right)\,\mu\,, &&                \label{denew}\\
\left(\gamma^{\mu\nu}\eta_\mu \eta_\nu + \tilde{\mu}^2\right)
\psi(\bp)=0\,,\quad{\mbox{with}}\;\;
\tilde{\mu}^2
= \left(
\begin{array}{cc}
-\zeta^{-1} \1_3 & 0_3 \\
0_3 & -\zeta \1_3
\end{array}
\right)\,\mu^2\,.&&
\label{bwweqnew}
\eeq
From an operational point of view the $\eta^\mu$ and $\mu$
\textit{are} the physical observables. The old operational meaning of
the symbols $E$ and $\p$ is lost in the non-linear realization
of the boost in the momentum space. There is
covariance of the form of the considered\footnote[4]{The result
is expected to be the same for other representation spaces.} fermionic 
and bosonic wave equations 
under the transformations:
\beq 
m \rightarrow \mu\,,\quad p^\mu \rightarrow \eta^\mu\,,\quad
\bv \rightarrow \bx\,. &&
\eeq
Thus, DSR and  special relativistic descriptions of fermions and bosons
become operationally indistinguishable.\\

\noindent
\textit{Towards a new relativity.}\textemdash~

The program of DSRs as presently formulated in terms of 
``phenomenologically motivated'' dispersion relations seems
untenable to us. Our reasons are as follows. In going from the
Galilean to special relativity it is true that one
needed a new scale. It was, as we know, the speed of light, $c$.
It unified space and time into one entity. Now if one wishes to introduce
another invariant scale, such as $\ell_p$, then the question arises as to
what aspect of spacetime it purports to unify, or
is there a new interaction which it brings in the realm
of spacetime structure. Recall that a 
new aspect in spacetime did emerge with the theory of general relativity
which endowed spacetime with the dynamics of 
the gravitational field.  With 
the work of Bekenstein and Hawking \cite{b,h,p}, one fused quantum and
thermodynamical features into the framework. Nothing similar seems to
have been
addressed in DSRs, and unless that happens we are afraid these suggestions
would remain devoid of any physical content.

Since the possible existence  of ultra high energy cosmic rays beyond the
GZK cutoff \cite{G,ZK} already point to a possible
violation of the Lorentz symmetry and emergence of new dispersion
relation[s] \cite{TLM,KM,T,DF,M,EMN}, it should be asked if there exists
a natural mechanism which asks for a new relativity. 
One such possibility
has unexpectedly emerged in a recent work on Majorana spinors 
\cite{hep-ph/0212222}. The essential idea, for
spin one half, is to  
promote the phase $\zeta$ to a complex $2\times 2$  
phase matrix with determinant $\pm 1$. 
Its precise form is then fixed 
by demanding that the resulting spinors have some
pre assigned  property under one of the discrete operators: C, P, and T.
Preliminary analysis of Ref. \cite{hep-ph/0212222} suggests that in
this process one
obtains new dispersion relations which demand specific modifications
in our conception of spacetime. 
Since such a global relative phase \textendash~ or, phase matrix
\textendash~  encodes C, P, and T properties of matter fields
without directly becoming source of energy momentum tensor,
the spacetime  
remains flat.  
But, now it incorporates the new quantum phase aspect into its
structure and leads to the identification of a preferred frame
in the context of Majorana particles
\cite{hep-ph/0212222,ADKP}. 
In contrast to DSRs,
in the considered framework 
there is a clear picture of what is purported to be synthesized 
\textendash~
the extension to special relativity is now sought in the same spirit as
was done to get a theory of gravitation in general relativity.
The new aspect emerges due to introduction of a C, P, and T encoding
phase field.
It appears, at present,
that in such a theory the Dirac particles are associated with
the standard dispersion relation while a new dispersion relation
emerges for the Majorana-like sector and takes the form:
$E=2m\pm\sqrt{\p^2+m^2}$.
Eventually, in presence of gravitation, 
the new phase field may have to be made local.

 \ack
One of us (DVA) thanks Giovanni Amelino-Camelia, 
Gaetano Lambiase, and their respective institutes
in Rome and Salerno, for discussions and hospitality
in May 2001. We also extend our thanks 
to Parampreet Singh for discussions. 

In part, this work is supported by Consejo Nacional de Ciencia y
Tecnolog\'ia (CONACyT, Mexico).

\end{document}